\documentclass[conference, letterpaper, 10pt]{IEEEtran}
\ifCLASSINFOpdf
\else
\fi
\usepackage{amsmath}
\usepackage{algorithmicx}
\usepackage{algorithm}
\usepackage{subcaption}
\usepackage{algcompatible}
\usepackage{graphicx}
\usepackage{amssymb}
\usepackage{epstopdf}
\usepackage[flushleft]{threeparttable}
\usepackage[keeplastbox]{flushend}
\usepackage{balance}
\usepackage{eso-pic}
\newcommand\blfootnote[1]{%
  \begingroup
  \renewcommand\thefootnote{}\footnote{#1}%
  \addtocounter{footnote}{-1}%
  \endgroup
}
\begin{document}
\bstctlcite{IEEEexample:BSTcontrol}
% Title
%\title{Design and Evalution of Rate-Maximizing Pilot Configurations for V2X Scenarios}
%\title{Self-Optimizing OFDM Pilot Configurations for Air-to-Ground UAV Communications}
\title{Rate-Maximizing OFDM Pilot Patterns for UAV Communications in Nonstationary A2G Channels}
\author{\IEEEauthorblockN{Raghunandan M. Rao\IEEEauthorrefmark{1}, Vuk Marojevic\IEEEauthorrefmark{1},
Jeffrey H. Reed\IEEEauthorrefmark{1} 
}
\IEEEauthorblockA{\IEEEauthorrefmark{1}Bradley Department of Electrical and Computer Engineering\\ Virginia Tech, Blacksburg, Virginia, USA\\ 
Email: \{raghumr, maroje, reedjh\}@vt.edu}
}

\maketitle
\begin{abstract}
In this paper, we propose and evaluate rate-maximizing pilot configurations for Unmanned Aerial Vehicle (UAV) communications employing OFDM waveforms. OFDM relies on pilot symbols for effective communications. We formulate a rate-maximization problem in which the pilot spacing (in the time-frequency resource grid) and power is varied as a function of the \textit{time-varying channel statistics}. The receiver solves this rate-maximization problem, and the optimal pilot spacing and power are \emph{explicitly fed back} to the transmitter to adapt to the time-varying channel statistics in an air-to-ground (A2G) environment. We show the enhanced throughput performance of this scheme for UAV communications in sub-6 GHz bands. These performance gains are achieved at the cost of very low computational complexity and feedback requirements, making it attractive for A2G UAV communications in 5G.

\textit{Index Terms}---UAV Communications, Rate Maximization, Adaptive Pilot Patterns, Channel Statistics Codebook.
\end{abstract}

\IEEEpeerreviewmaketitle
%Standard footnote as in all prior ArXiv uploads
\blfootnote{This is the author's version of the work. For citation purposes, the definitive version of record of this work is: R. M. Rao, V. Marojevic and J. H. Reed, ``Rate-Maximizing OFDM Pilot Patterns for UAV Communications in Nonstationary A2G Channels,'' \textit{To Appear in the 88th IEEE Vehicular Technology Conference (IEEE VTC Fall 2018)}, pp. 1-- 5, August 2018.}

\section{Introduction}\label{Sec1_Intro}
Unmanned Aerial Vehicle (UAV) communications has spurred a lot of interest in recent years, in particular due to an upsurge in its wide-ranging applications such as Internet by drones, package delivery and public safety networks. The Third Generation Partnership Project (3GPP) has frozen Release 15 in December 2017, which publishes the first specifications for 5G New Radio (5G NR). Integrating UAVs with 5G is one of the most promising approaches being considered to facilitate ubiquitous high speed as well as low latency communications.

%However, the expected densification of UAV communication networks will challenge the limits of 5G 
Maintaining a high spectral efficiency in the face of impending densification of UAVs will be challenging in the future, particularly in the crowded sub-6 GHz bands. Moreover, the channel statistics of air-to-ground channels will be significantly different than those of terrestrial wireless channels. For instance, UAV communications operate in a Line of Sight (LoS) or near-LoS environments, unlike the rich scattering multipath environments observed in terrestrial channels. Moreover, high mobility UAV scenarios such as package delivery, mission-specific military drones etc. will experience fast temporal fading due to Doppler shifts. Since 5G is anticipated to encompass a wide variety of channel scenarios, it is more efficient to allow for self-optimization in order to adapt to time-varying channel environments.

Xue et al. \cite{Xue_TF_Sched_UAV_2018} present joint time-frequency scheduling and power allocation schemes to manage the effect of channel fading and adjacent channel interference in multi-UAV communications. He et al. \cite{He_Alt_BeamW_UAV_2018} optimize the UAV's altitude and antenna beamwidth, coupled with a fly-hover-and-communicate protocol to efficiently serve ground terminals partitioned into disjoint clusters. Jaber et al. \cite{Jaber_Vuk_UAS_2017} optimize OFDM parameters as a function of carrier frequency and UAV speed or type.

\begin{figure}[t]
\centering
\includegraphics[width=3.3in]{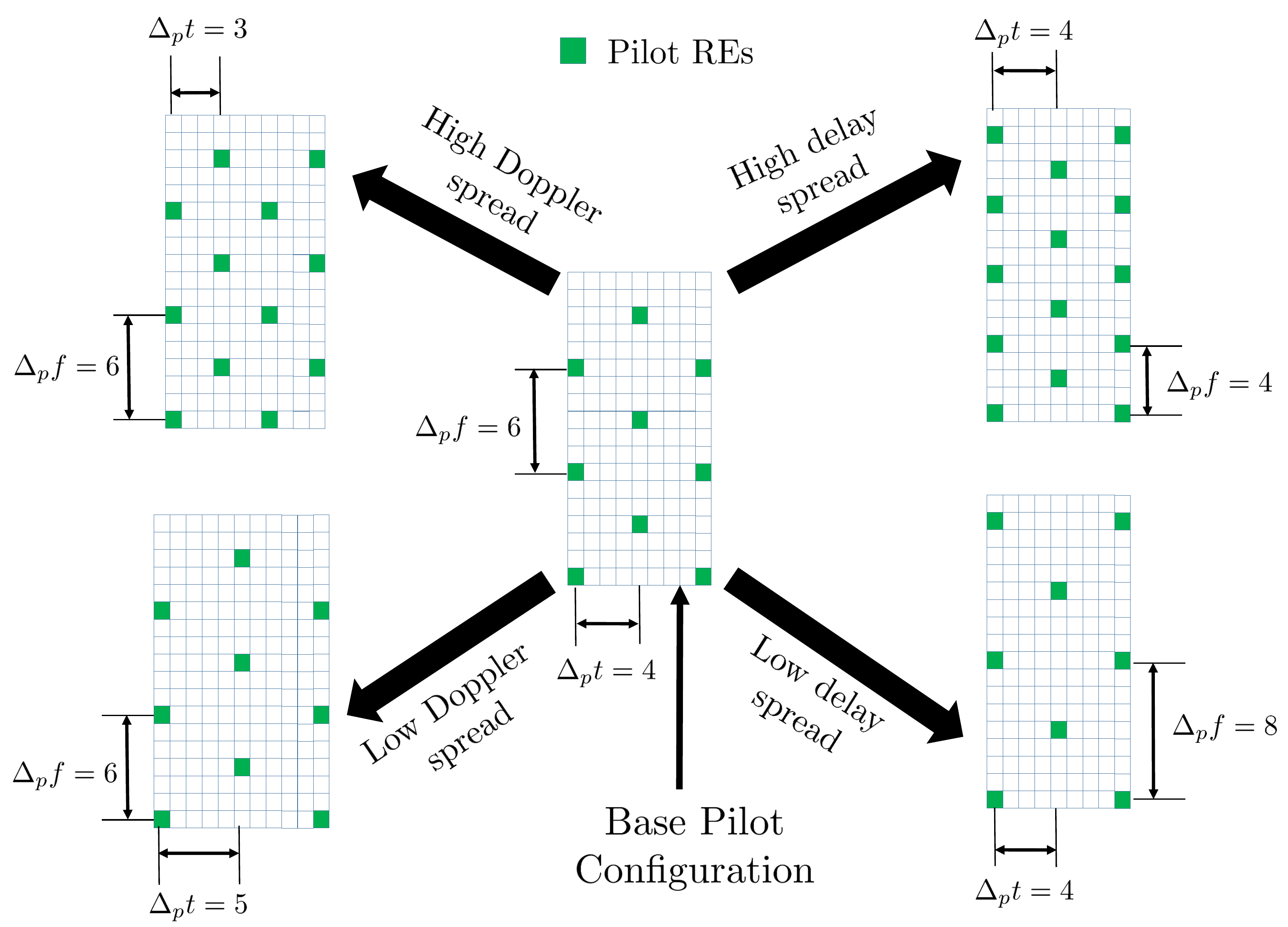}
\caption{Illustration of pilot parameter adaptation in the OFDM resource grid as a function of channel statistics.}
\label{Fig1_pilot_adapt_illustrate}
\end{figure}

In modern wireless standards, pilots are used for coherent demodulation and channel state information (CSI) estimation. Current wireless standards employ fixed pilot configurations that are designed for the worst-case channel conditions. With 4G LTE, the pilot pattern was primarily designed for terrestrial communications. However, this is clearly not the case for UAV communications in LoS and potentially fast fading A2G channels. Fig. \ref{Fig1_pilot_adapt_illustrate} illustrates how the pilot spacing can be changed as a function of the Doppler and delay spreads to balance pilot overhead with channel estimation accuracy.  

Whereas most of the recent literature has focused on nearly static channel conditions or low-speed UAVs, we develop a methodology for optimizing pilot signal configurations to maximize rate for UAVs in A2G channels with time-varying statistics, specifically, Doppler spread and delay spread. As \cite{simko2013adaptive} and \cite{Raghu_TVT_2017} have shown, the rate-maximizing pilot pattern is a function of the time-frequency fading characteristics, and the operating SNR of the channel. Although 5G NR allows varying the pilot density in the time domain, (a) it has a limited number of configurations, and (b) it does not allow changing pilot density in the frequency domain  \cite{phy_channels_modulation}, which limits resources for data in LoS A2G channels. 

%In this paper, we design and evaluate adaptive rate-maximizing pilot patterns for UAV communications in an A2G channel with a time-varying SNR, Doppler spread and delay spread. %The channel model we've used is based on real measurements reported in \cite{Matolak_AG_Hill_2017}-\cite{Matolak_AG_Urb_2017}. 
Our approach follows the principles of \cite{Raghu_TVT_2017}. We compare the throughput performance of our proposed rate-maximizing pilot scheme against other fixed pilot configurations in a \textit{realistic A2G channel with time-varying statistics}. We also explore explicit feedback mechanisms to facilitate dynamic pilot adaptation. This is opposed to implicit feedback, which was introduced in \cite{Raghu_TVT_2017}. We observe that signaling overhead for explicit feedback is negligible, and is of the same order of magnitude as implicit feedback, with low computational complexity. %Its reduction of computational complexity at the transmitter, coupled with the negligible feedback overhead makes it attractive for 5G.  

The rest of the paper is organized as follows: Section \ref{Sec2_ProbForm} formulates the optimization problem for finding rate-maximizing pilot patterns. Section \ref{Sec3_ChanStatCB} introduces the channel statistics codebook, and discusses the feedback mechanisms to enable pilot adaptation. Section \ref{Sec4_NumResults} provides numerical results and analysis comparing our adaptive rate-maximizing pilot design with that of fixed pilot configurations. Section \ref{Sec5_Conc} concludes the paper. 

\section{Problem Formuulation and System Model}\label{Sec2_ProbForm}

%More recently, 5G allows  varying the pilot density, the number of options for varying the pilot spacing  in time [5GSpecRef]. 
\subsection{Rate-Maximizing Pilot Configurations}
Finding the optimal pilot configuration can be formulated as a maximization problem of the upper bound of the achievable rate \cite{simko2013adaptive}, \cite{Raghu_TVT_2017}, which can be written as
\begin{align}
\label{optimization_problem}
\underset{ \{\rho, \Delta_p f, \Delta_p t \} } {\text{maximize}} \quad &  S(\Delta_p f, \Delta_p t)\cdot \log_2 (1 + \bar{\gamma}) \\
\text{subject to} \quad & \bar{P}(\rho, \Delta_p f, \Delta_p t) \leq 1 \nonumber \\
\quad & 1 \leq \Delta_p t \leq T_{max} \nonumber \\
\quad & 2 \leq \Delta_p f \leq F_{max} \nonumber \\
\quad & \rho_{min} \leq \rho \leq \rho_{max}, \nonumber
\end{align}
where $\Delta_p t$ and $\Delta_p f$ are the pilot spacing in time and frequency. The average data-to-pilot power ratio is given by $\rho=\sigma_d^2/\sigma_p^2$. %, where $\sigma_d^2$ is the average power per data symbol, and $\sigma_p^2$ that for the pilot symbols. 
The triad $\mathcal{V}=\{\rho, \Delta_p f, \Delta_p t \}$ completely describes the pilot configuration. %, which is varied as a function of the channel statistics as shown in Fig. \ref{Fig1_pilot_adapt_illustrate}. 
The upper limits on pilot spacing in time and frequency is given by $T_{max}$ and $F_{max}$ respectively, which is found using sampling considerations \cite{ChanEstOFDM_Param_model}. The average power per resource element is given by $\bar{P}(\mathcal{V})$, which is a function of the pilot configuration as shown in \cite{Raghu_TVT_2017}. The upper (lower) limit on $\rho$ is denoted by $\rho_{max}\ (\rho_{min})$ respectively. The average post-equalization SINR $(\bar{\gamma})$ for a zero-forcing (ZF) receiver can be written as
\begin{equation}
\label{PostSqSINR_ZF}
\bar{\gamma} = \frac{\sigma_d^2 \cdot \sigma_{ZF}}{\sigma_w^2 + \sigma_{ICI}^2 + \sigma_{MSE}^2 \cdot \sigma_d^2}, 
\end{equation}
where $\sigma_d^2\ (\sigma_p^2)$ is the average power per data (pilot) symbol, $\sigma_w^2$ the noise power and $\sigma_{ZF} = 1$ for a $M\times M\text{-MIMO}$ system \cite{simko2013adaptive}. The channel estimation mean squared error (MSE) is given by $\sigma_{MSE}^2$, which can be computed using the expressions derived in \cite{simko2013adaptive}, \cite{Raghu_TVT_2017}\footnote{The channel estimation MSE has been derived for `diamond-shaped' pilot patterns in \cite{simko2013adaptive}, \cite{Raghu_TVT_2017}. Note that this pattern is used in modern cellular standards such as LTE and NR. $\sigma_{MSE}^2$ for other frequency/time comb patterns can be derived in a similar manner.}. It is important to note that $\sigma_{MSE}^2$ is a function of the channel's temporal correlation $R_t (\Delta t)$, spectral correlation $R_f(\Delta f)$ and the average pilot SNR ($\sigma_p^2/\sigma_w^2$). The intercarrier interference $\sigma_{ICI}^2$ is assumed to be dominated by user mobility in the vehicular network and can be estimated using \cite{ICI_bounds_Cimini_2001}

\begin{equation}
\sigma_{ICI}^2 \leq \frac{1}{3} \Big(\frac{\pi f_d \sigma_d}{\Delta f} \Big)^2,
\end{equation}
where $f_d$ is the maximum Doppler shift and $\Delta f$ the subcarrier spacing. The spectrum utilization function $S(\Delta_p f, \Delta_p t)$ is simply the fraction of data REs in the OFDM grid across all layers over the total number of REs, and can be easily computed with the knowledge of $\mathcal{V}$. Additional control channels and signals are ignored here without loss of generality. In order to estimate the achievable rate as a function of $\mathcal{V}$, the unknown quantities that need to be estimated are $\sigma_{MSE}^2$, $f_d$ and $\sigma_w^2$.

Noise power can be estimated using the methods proposed in \cite{Cui_PDP_NoiseVarEst_2006}. To estimate the channel statistics $\hat{R}_t (\Delta t)$ and $\hat{R}_f (\Delta f)$ in a nonstationary vehicular environment, temporal averaging can be performed assuming local stationarity \cite{Raghu_TVT_2017}, \cite{Matolak_AG_Urb_2017}, \cite{Matolak_AG_Hill_2017}. With the estimated channel matrix $\hat{\mathbf{H}} \in \mathbb{C}^{N_{sub} \times T_{ofdm}}$, the channel correlation can be estimated using
\begin{align}
\label{estimate_statistics}
\hat{R}_t (-i) &= \frac{1}{T_{ofdm} - \lvert i \rvert} \sum_{t = 1}^{T_{ofdm} - \lvert i \rvert} \Big \{\text{diag}_i \Big[ \hat{\mathbf{H}}^H \hat{\mathbf{H}}  \Big] \Big \}_t  \nonumber \\
\hat{R}_f (-j) &= \frac{1}{N_{sub} - \lvert j \rvert} \sum_{f = 1}^{N - \lvert j \rvert} \Big \{ \text{diag}_j  \Big[ \hat{\mathbf{H}} \hat{\mathbf{H}}^H  \Big] \Big \}_f ,
\end{align}
where $N_{sub}$ is the number of subcarriers and $T_{ofdm}$ the number of OFDM symbols in the channel statistics estimation window. The term $\text{diag}_i [\mathbf{X}]$ is the vectorized $i^{th}$ diagonal of matrix $\mathbf{X}$, and $\Big \{\text{diag}_i [\mathbf{X}] \Big \}_k$ its $k^{th}$ element. Due to conjugate symmetry, the other elements can be found using $\hat{R}_t (-i) = \hat{R}^*_t (i)$ and $\hat{R}_f (-j) = \hat{R}^*_f (j)$. This formulation can be extended to MIMO-OFDM, where the channel spectral and temporal correlation is estimated for each layer\footnote{Typically the spectral and temporal correlation is the same for the channel between each tx-rx antenna pair, unless the antennas are distributed in different locations of the network.}.

\section{Practical Channel Statistics Estimation and Feedback}\label{Sec3_ChanStatCB}
In practical scenarios where the channel statistics are estimated over a finite duration, the accuracy will degrade due to (a) interpolation errors, and (b) addition of noise. In the worst case, the estimated channel statistics can violate the properties of the autocorrelation function $\vert \hat{R}_t (\Delta t) \rvert \leq \hat{R}_t (0)\ \forall\ \Delta t \neq 0 $. This can happen especially in high noise, low mobility and/or flat fading scenarios. Using these estimated channel statistics directly can result in inconsistent, and sometimes absurd values for the MSE. We propose a codebook-based approach to mitigate this issue, as well as reduce feedback requirements. %The codebook contains the power delay profile (PDP) and maximum Doppler frequency values of typical channels that the radio expects to encounter. A cognitive radio, for example, can update the codebook over time as it learns more about its channel environment. The receiver calculates the channel statistics using equation (\ref{estimate_statistics}) for a finite duration and finds the codebook profile that is closest to it in the minimum euclidean distance sense.
\subsection{Channel Statistics Codebook}
We propose a codebook that contains a finite number of channel statistics, i.e. channel correlations in the time and frequency dimensions. Let the codebook be denoted by set $\mathcal{R}_C$ with two sets $\mathcal{R}_{C,t}\in \mathcal{R}_C$ and $\mathcal{R}_{C,f} \in \mathcal{R}_C$. Let $\lvert \mathcal{R}_{C,f} \rvert = M_f$ and $\lvert \mathcal{R}_{C,t} \rvert = M_t$, where $\lvert \cdot \rvert$ denotes the cardinality of a set. $\mathcal{R}_{C,f}$ is the set of channel frequency correlation profiles with vector elements $\mathbf{R_{fc,l}} \in \mathcal{R}_{C,f}, 1 \leq l \leq M_f$. Likewise, $\mathcal{R}_{C,t}$ is the set of channel temporal correlation profiles with vector elements $\mathbf{R_{tc,m}} \in \mathcal{R}_{C,t}, 1 \leq m \leq M_t$. Here, we model temporal fading using a classic Doppler spectrum where the $(\Delta t)^{th}$ element of $\mathbf{R_{tc,m}}$ is $[\mathbf{R_{tc,m}}]_{\Delta t} = J_0(2 \pi f_{d,m} \Delta t)$, with $f_{d,m}$ being the maximum Doppler frequency for the $m^{th}$ temporal correlation profile. This codebook design is motivated by the Wide sense stationary uncorrelated scattering (WSSUS) approximation, which allows for independent modeling of multipath fading and user mobility. In general, codebook elements of $\mathcal{R}_{C,f}\ (\mathcal{R}_{C,t})$ is a vector of length $N_{\Delta t}\ (N_{\Delta f})$ respectively. The vector lengths $N_{\Delta t}$ and $N_{\Delta f}$ must be chosen to balance accuracy of channel correlation estimation and computational complexity. For the sake of representation $\mathcal{R}_{C,t}$ ($\mathcal{R}_{C,f}$) can be parametrized by $f_d$ ($\tau_{rms}$) respectively, as shown in Table \ref{Tab2_Chan_Stat_Codebook}. 
%Initially, the profiles that comprise the codebook would correspond to the most common types of channels that the radio would be expected to encounter, based on reported field measurements. For example the channel profiles from ITU-T \cite{ITUIMT2000spec} and the 3GPP channel models \cite{3GPPLTE_TS36_104_tx_and_rx} can be used as initial codebook entries. In the case of a cognitive radio, the codebook can be updated over time, when it learns more about its operating channel environment. The codebook can be designed to match the typical scenarios operation environment of the radios. For example vehicular to vehicular networks would have a large variation in Doppler spreads. On the other hand, UAV-to-UAV systems might have very low root mean square delay spread due to strong line of sight propagation \cite{Malotak_UAVCHanChar_2016, Zeng_UAV_Chall_2016}. We will provide example codebooks in the next section.
\subsection{Estimation and Feedback of Optimal Parameters}
\begin{figure}[t]
\centering
\includegraphics[width=3.3in]{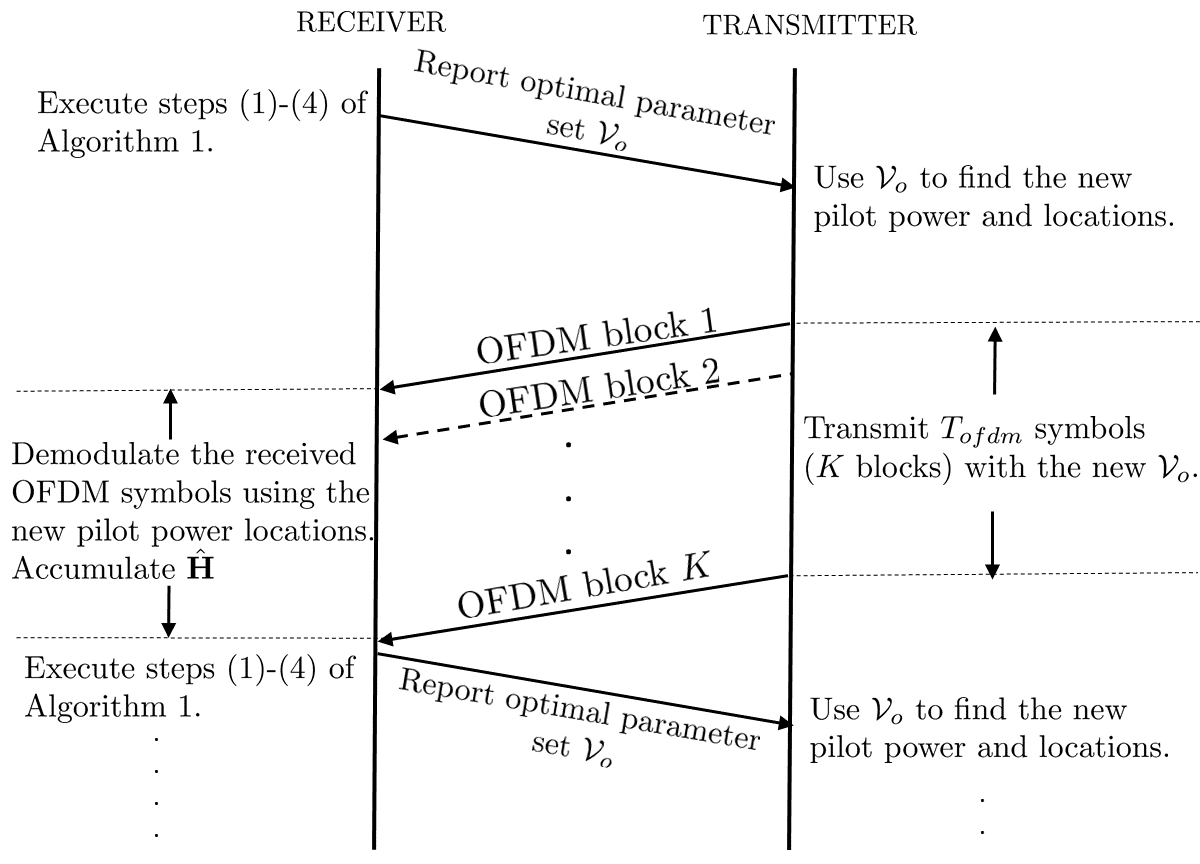}
\caption{Illustration of the explicit feedback of pilot parameters between the transmitter and the receiver based on Algorithm \ref{algo_pilot_adapt}. $K$ OFDM blocks are equivalent to $T_{ofdm}$ OFDM symbols.}
\label{Fig_Pil_adapt_algo_exchange}
\end{figure}

In order to reliably communicate, the transmitter and receiver should share $\{\mathcal{P}, \mathcal{D}_f, \mathcal{D}_t\}$, which are the sets for $\{\rho, \Delta_p f, \Delta_p t\}$ respectively. Using (\ref{optimization_problem}), Algorithm \ref{algo_pilot_adapt} finds $\mathcal{V}_o$ and updates it every $T_{ofdm}$ symbols. For small discrete-valued feasible sets, a simple brute force method to find $ \mathcal{V}_o$ is practically feasible. The receiver feeds back $\mathcal{V}_o$, which is then used by the transmitter for transmission for the next $T_{ofdm}$ OFDM symbols. Fig. \ref{Fig_Pil_adapt_algo_exchange} shows the processing and feedback of codebook indices for pilot adaptation. The computational complexity of Algorithm \ref{algo_pilot_adapt} is low since it is composed of matrix multiplication operations and optimization problems involving discrete-valued, low-dimensional feasible sets.

\begin{algorithm}[t]
\small
\begin{algorithmic}[1]
\STATEx \textbf{Input:} \emph{Codebook $\mathcal{R}_C$} and $\{ \mathcal{D}_f, \mathcal{D}_t \text{ and } \mathcal{P} \}$.

\STATE Estimate $\hat{R}_t$ and $\hat{R}_f$ from equation (\ref{estimate_statistics}) using $\mathbf{\hat{H}}$, computed using the most recent $T_{ofdm}$ OFDM symbols. 

\STATE Find $\mathbf{R_{fc, l'}} \in \mathcal{R}_{C,f}$ and $\mathbf{R_{tc, m'}} \in \mathcal{R}_{C,t}$ by solving
\begin{align}
\label{optimize_pilots}
l' = \underset{1 \leq l \leq M_f} {\text{arg min}} \quad
& \| \mathbf{\hat{R}_f} - \mathbf{R_{fc, l}} \| \nonumber \\
m' = \underset{1 \leq m \leq M_t} {\text{arg min}} \quad
& \| \mathbf{\hat{R}_t} - \mathbf{R_{tc, m}} \|. 
\end{align}
%\noindent For a $N_{tx} \times N_{rx}$ MIMO-OFDM, there will be $N_{tx} N_{rx}$ channel matrices of dimension $N \times T_{ofdm}$ (one for each transmit-receive antenna pair). If $\mathbf{l'} \text{ and } \mathbf{m'}$ represent the $N_{tx} N_{rx} \times 1$ vectors of codebook indices found using equation (\ref{optimize_pilots}) for each channel matrix, then $l' = \text{mode}(\mathbf{l'}), m' = \text{mode}(\mathbf{m'})$. %Here, $\text{mode}(.)$ represents the `mode' operation.  

%\STATE Feed back the codebook indices $l' \text{ and } m'$ to the transmitter on the uplink.

\STATE For all allowed values of $\mathcal{V}=\{\rho, \Delta_p f, \Delta_p t\} \in \{\mathcal{P}, \mathcal{D}_f, \mathcal{D}_t\}$, compute $\sigma_{MSE}^2$ (see \cite{Raghu_TVT_2017}).

\STATE For all allowed values of $\mathcal{V}$, solve equation (\ref{optimize_pilots}) to obtain the optimal parameters $\mathcal{V}_o=\{ \rho_{o},(\Delta_p f)_{o},(\Delta_p t)_{o}\} $.

\STATE Feed back the optimal parameter set $\mathcal{V}_o$.

\STATE For the next $T_{ofdm}$ OFDM symbols, use $\mathcal{V}_o$ to estimate the new channel matrix $\mathbf{\hat{H}} \in \mathbb{C}^{N_{sub} \times T_{ofdm}}$. 

\STATE Go back to step 1.
\end{algorithmic}
\caption{Pilot Adaptation using Explicit Feedback: Receiver Processing and Signaling}
\label{algo_pilot_adapt}
\end{algorithm}

\subsection{Feedback Mechanisms and Requirements}
Here we discuss two methods for feeding back the optimal pilot configuration.
\subsubsection{Explicit Feedback of the Optimal Parameters}
The receiver can use explicit feedback of the the optimal parameters $\mathcal{V}_o$ to facilitate pilot adaptation as shown in Fig. \ref{Fig_Pil_adapt_algo_exchange}. In this case, the minimum number of bits required will be $b_{\mathtt{exp}}^{(fb)} = \lceil \log_2 (M_{\mathcal{P}} M_{\mathcal{D}_f} M_{\mathcal{D}_t}) \rceil$ bits, where $M_{\mathcal{P}} = \lvert \mathcal{P}\rvert$, $M_{\mathcal{D}_f} = \lvert \mathcal{D}_f \rvert$ and $M_{\mathcal{D}_t} = \lvert \mathcal{D}_t \rvert$. Since the codebook indices are fed back once every $(T_{ofdm}T_s)$ seconds, the rate overhead for explicit feedback will be $R_{\mathtt{exp}}^{(fb)} = b_{\mathtt{exp}}^{(fb)}/(T_{ofdm} T_s) \text{ bps}$. %$R_{\mathtt{exp}}^{(fb)} = \tfrac{b_{\mathtt{exp}}^{(fb)}}{T_{ofdm} T_s} \text{ bps}$. 
 %since the channel statistics (doppler and r.m.s. delay spread) vary slowly in typical vehicular scenarios
\subsubsection{Implicit Feedback using Codebook Indices}
If the cardinality of the feasible set is large, the feedback requirements can be further reduced by implicit feedback, where the codebook indices $(l', m')$ are fed back instead of $\mathcal{V}_o$. 
In this case, the minimum number of bits required is  $b_{\mathtt{imp}}^{(fb)} = \lceil \log_2(M_t M_f) \rceil$ bits. The rate overhead for implicit feedback will be $R_{\mathtt{imp}}^{(fb)} = b_{\mathtt{imp}}^{(fb)}/(T_{ofdm} T_s) \text{ bps}$. %For more details about implicit feedback, the interested reader is referred to \cite{Raghu_TVT_2017}. %$R_{\mathtt{imp}}^{(fb)} = \tfrac{b_{\mathtt{imp}}^{(fb)}}{T_{ofdm} T_s} \text{ bps}$. 
%As we will discuss in our results, the additional feedback overhead for both implicit and explicit feedback for pilot adaptation will be negligible in current wireless standards.
\begin{figure*}[t]
    \centering
    \begin{subfigure}[t]{0.3\textwidth}
    \label{3a}
        \raggedleft
        \includegraphics[width=1.9in]{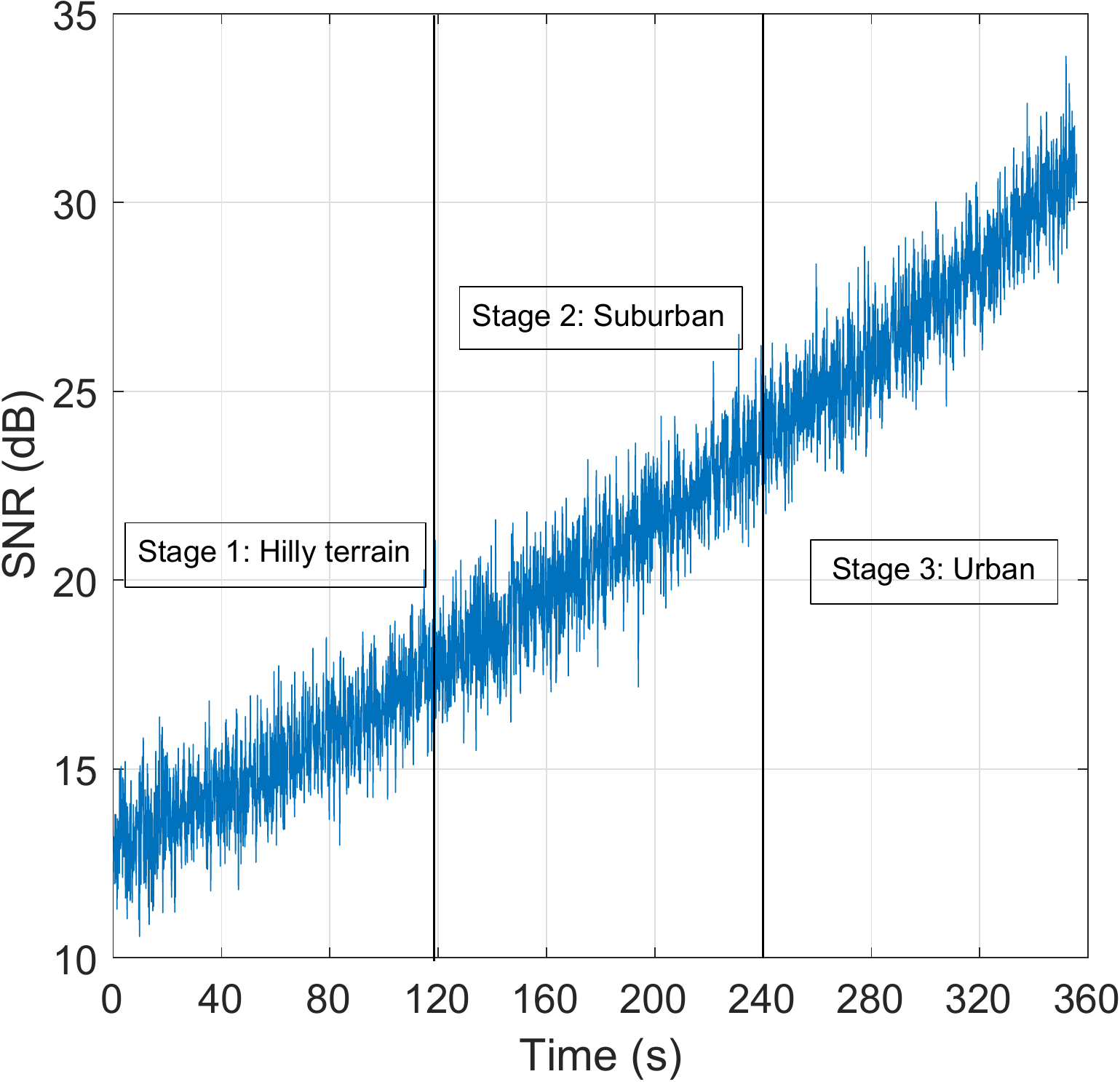}
        \caption{}
    \end{subfigure}%
    ~ 
    \begin{subfigure}[t]{0.3\textwidth}
    \label{3b}
        \centering
        \includegraphics[width=1.9in]{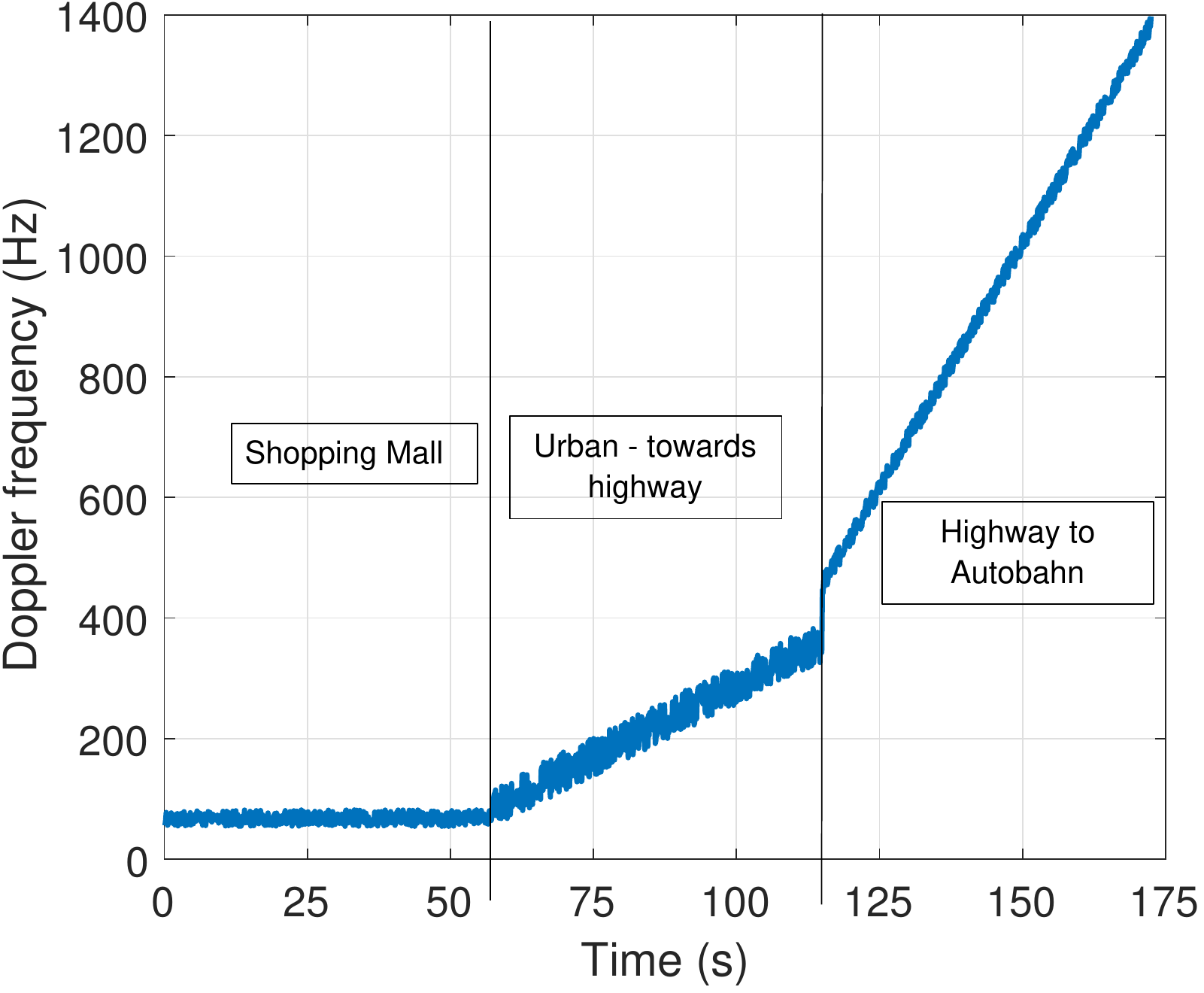}
        \caption{}
    \end{subfigure}
    ~
    \begin{subfigure}[t]{0.3\textwidth}
    \label{3c}
        \centering
        \includegraphics[width=1.9in]{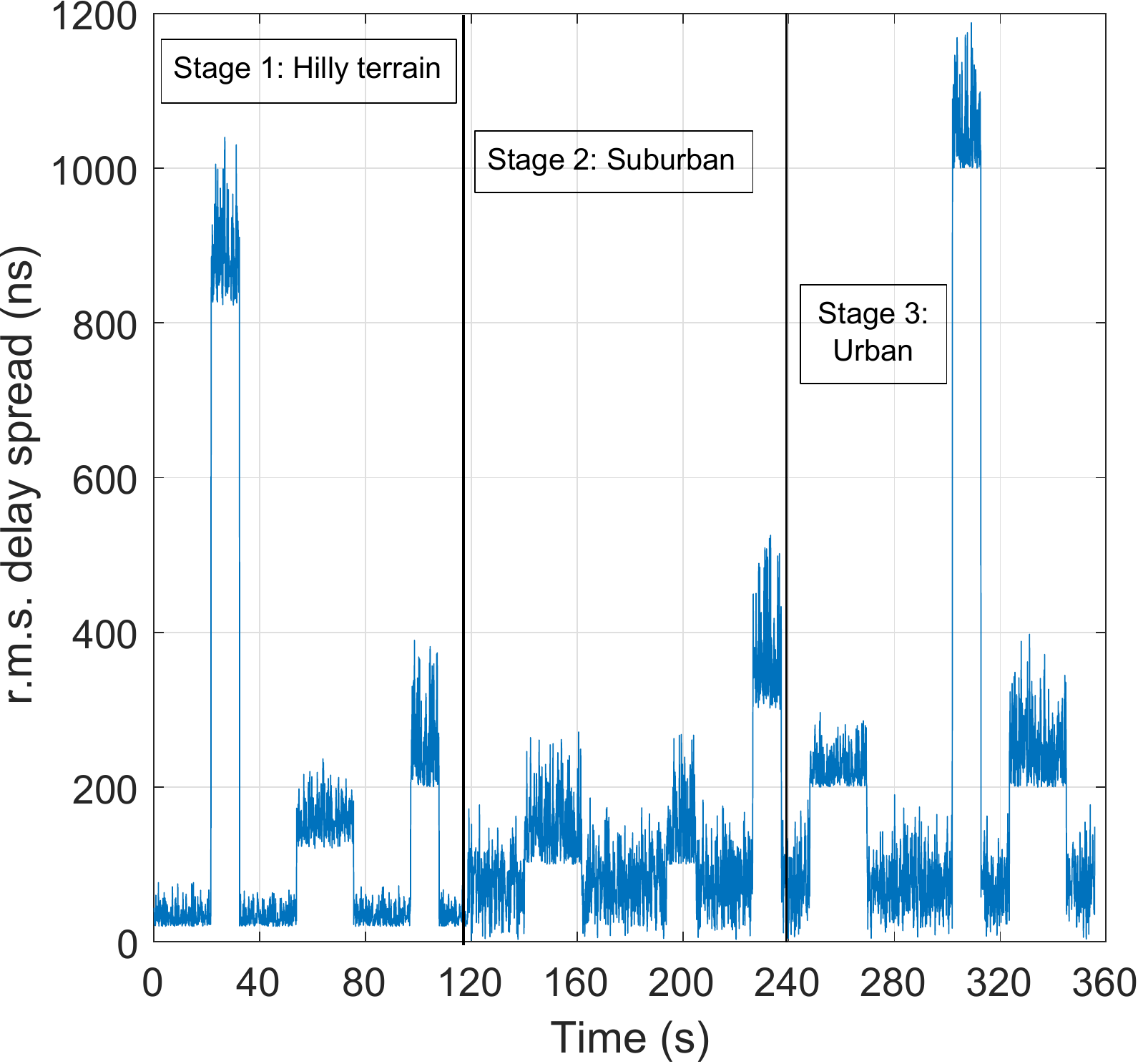}
        \caption{}
    \end{subfigure}
    \caption{Variation of (a) SNR, (b) $f_d$ and (c) $\tau_{rms}$ over time in the simulation scenario. }
    \label{Fig3_scenario}
\end{figure*}

\section{Numerical Results}\label{Sec4_NumResults}
%In this section, we will show the capacity gain achievable due to pilot adaptation based on the channel fading statistics. 
We consider a A2G wireless channel in the 5 GHz band. The wireless channel can be parametrized by the signal-to-noise ratio (SNR), Doppler spread ($f_d$) and the root-mean squared (r.m.s.) delay spread ($\tau_{rms}$). To model the wireless channel, we used the tapped-delay line model with a time-varying power delay profile (PDP) and the Jakes Doppler spectrum applied on each multipath component with the appropriate $f_d$.

Although a channel with time varying statistics is nonstationary, the channel can be approximated to be stationary within a distance called the stationarity distance (SD). For A2G channels, extensive channel measurements in \cite{Matolak_AG_Urb_2017}, \cite{Matolak_AG_Hill_2017} have shown that the SD ranges between $10 \text{ and } 35 \text{ m}$. For a UAV traveling at an average speed of $75 \text{ m/s}$ this corresponds to a stationarity time of up to $450 \text{ ms}$. %We consider a vehicular channel with slowly varying statistics, where $f_d$ and $\tau_{rms}$ change on the order of hundreds of milliseconds, as reported in \cite{Bernado_vehic_nonstat_2014}. The simulation parameters are shown in Table \ref{Tab1_sim_parameters}. 

\begin{table}[t]
%% increase table row spacing, adjust to taste
\renewcommand{\arraystretch}{1.0}
% if using array.sty, it might be a good idea to tweak the value of
% \extrarowheight as needed to properly center the text within the cells
\caption{Simulation Parameters}
\label{Tab1_sim_parameters}
\centering
%% Some packages, such as MDW tools, offer better commands for making tables
%% than the plain LaTeX2e tabular which is used here.
\begin{tabular}{|l|l|}
\hline
\textbf{Parameter} & \textbf{Value} \\
\hline
Antenna Configuration & SISO \\
\hline
FFT-length & 128\\
\hline
No. of subcarriers $(N_{sub})$ & 72 \\
%\hline
%Number of Guard Subcarriers & 28 on each band edge\\
\hline
Center Frequency $(f_c)$ & $5 \text{ GHz}$ \\
\hline
Subcarrier Spacing $(\Delta f)$ & $15\ \text{kHz}$\\
\hline
OFDM symbol duration $(T_s)$ & $71.875\ \mu s$\\
%$T_{OFDM}$ 
\hline
Cyclic Prefix Duration & $5.21\ \mu s$\\
\hline
Base pilot spacing & $\Delta_p t =  4\ (0.2875 \text{ ms})$ \\
& $\Delta_p f = 6\ (90 \text{ kHz})$ \\
\hline
Channel parameters & Doubly selective: Jakes Doppler \\
 & spectrum with multipath fading.\\
\hline
Transmit power & $37.5 \text{ dBm}$ \\
\hline
Noise Power Spectral Density & $-174 \text{ dBm/Hz}$\\
\hline
Pathloss parameters \cite{Matolak_AG_Urb_2017} & $A = 116 \text{ dB}$, $n = 1.8$, $\sigma_X = 3.1 \text{ dB}$\\ 
& $F = 2.3 \text{ dB}$, $R_{max} = 19 \text{ km}$ \\
& $R_{min} = 1.7 \text{ km}$  \\
\hline
Channel Estimation & Least Squares (pilots) \\
& 2D-Linear Interpolation (data REs)\\
\hline
Equalization & Zero Forcing (ZF) \\
\hline
\end{tabular}
\end{table}

\subsection{Scenario}
We consider a scenario where a UAV is communicating with a ground station (GS) using an OFDM (LTE or NR-like) PHY layer (Table \ref{Tab1_sim_parameters}). %Fig. \ref{Fig3_scenario} shows the variation of the channel parameters with time. The simulation parameters are shown in Table \ref{Tab1_sim_parameters}. 
The scenario consists of three stages, each lasting for about 2 minutes:
\begin{itemize}
\item Stage 1: The UAV flies in a hilly section towards a city. Due to reflections from hills, there is a presence of strong multipath components ($\tau_{rms} \sim 1 \mu s$ \cite{Matolak_AG_Hill_2017}), and the UAV decelerates from $300 \text{ km/h}$ to $200 \text{ km/h}$. 

\item Stage 2: The UAV then enters the suburban section, where $\tau_{rms}$ fluctuates between $50 \text{ ns}$ and $500\text{ ns}$ \cite{Matolak_AG_Urb_2017}. The UAV uniformly decelerates from $200 \text{ km/h}$ to $100 \text{ km/h}$.

\item Stage 3: The UAV enters the urban section, where the contributions of multipath become prominent due to the presence of tall buildings \cite{Matolak_AG_Urb_2017}. The UAV velocity decelerates further to $50 \text{ km/hr}$.
\end{itemize}

Regulations or UAV mission may be the cause for the varying UAV speeds, for e.g. a package delivery mission. The maximum doppler frequency $f_{d,m}$ is related to the velocity $v$ by $f_{d,m} = vf_c/c$, where $c$ is the velocity of light and $f_c$ the carrier frequency. The SNR varies with distance based on the pathloss model parameters shown in Table \ref{Tab1_sim_parameters}, using the distance-based path loss with log-Normal shadow fading \cite{Matolak_AG_Urb_2017} 
\begin{align}
\label{pathloss}
PL(d) & = A + 10n \log (d/R_{min}) + X - F \ \ \text{[dB]},
\end{align}
where $R_{min} \leq d \leq R_{max}$ and $X [\text{dB}] \sim \mathcal{N}(0, \sigma_X^2)$. 
%We consider a V2I/V2P link between a moving vehicle and almost immobile pedestrians (typical velocity of $3 \text{ km/hr}$) and CV2X infrastructure. The simulation scenario is shown in Fig. \ref{Fig3_scenario}, where a vehicle 
\begin{table}[!t]
%% increase table row spacing, adjust to taste
\renewcommand{\arraystretch}{1}
% if using array.sty, it might be a good idea to tweak the value of
% \extrarowheight as needed to properly center the text within the cells
\caption{Codebook of Channel Profiles, $\mathcal{R}_C$}
\label{Tab2_Chan_Stat_Codebook}
\centering
%% Some packages, such as MDW tools, offer better commands for making tables
%% than the plain LaTeX2e tabular which is used here.
\subcaption*{$\mathcal{R}_{C,t}$: Channel profiles for Doppler Frequency }
\begin{tabular}{|c|l|l|l|}
\hline
Index $(m)$ & Mobility Type & Velocity & $f_d^\dagger$ (Hz)\\
%& & & \\
\hline
1 & Almost stationary & $1 \text{ km/h}$ & 4.6\\
\hline
2 & Low speed (taxiing) & $15 \text{ km/h}$ & 70\\
\hline
3 & High speed (taxiing) & $55 \text{ km/h}$ & 250\\
\hline
4 & Takeoff/Landing & $120 \text{ km/h}$ & 550\\
\hline
5 & Medium speed (airborne) & $160 \text{ km/h}$ & 750\\ 
\hline
6 & High Speed (airborne) & $250 \text{ km/h}$ & 1150 \\
\hline
\end{tabular}
\subcaption*{$^\dagger$Doppler frequency for a center frequency of $f_c=5 \text{ GHz}$. }
\subcaption*{ $\mathcal{R}_{C,f}$: Channel profiles for Power Delay Profiles (PDP)}
\begin{tabular}{|c|c|c|}
\hline
Index $(l)$ & Type of Scattering & $\tau_{rms}$ \\
% & & (ns)\\
\hline
1 & Low (near-LoS) & 221.5\\
\hline
2 & Medium (Suburban air-to-ground) & 476.4\\
\hline 
3 & High (Near-Urban air-to-ground) & 791.2\\
\hline
4 & Very High (Urban/Hilly air-to-ground) & 1440\\
\hline
\end{tabular}
\end{table}

\subsection{Performance Comparison with Fixed Pilot Configurations}
We compare our rate-maximizing pilot scheme to the fixed pilot configurations $\mathcal{V}_{2,2}, \mathcal{V}_{4,2}, \mathcal{V}_{6,4}$ (similar to LTE), $\mathcal{V}_{6,6} \text{ and } \mathcal{V}_{8,8}$, where $\mathcal{V}_{a, b} = \{\rho, \Delta_p f, \Delta_p t \} = \{-3 \text{ dB}, a, b \}$.
%\begin{enumerate}
%\item $\mathcal{V}_{2,2} = \{\rho, \Delta_p f, \Delta_p t \} = \{-3 \text{ dB}, 2, 2 \}$.
%\item $\mathcal{V}_{2,4} = \{\rho, \Delta_p f, \Delta_p t \} = \{-3 \text{ dB}, 2, 4 \}$.
%\item $\mathcal{V}_{4,4} = \{\rho, \Delta_p f, \Delta_p t \} = \{-3 \text{ dB}, 4, 4 \}$ (similar to LTE's demodulation reference signal).
%\item $\mathcal{V}_{4,6} = \{\rho, \Delta_p f, \Delta_p t \} = \{-3 \text{ dB}, 4, 6 \}$ (similar to LTE's cell-specific reference signal).
%\item $\mathcal{V}_{8,8} = \{\rho, \Delta_p f, \Delta_p t \} = \{-3 \text{ dB}, 8, 8 \}$.
%\end{enumerate}
Table \ref{Tab2_Chan_Stat_Codebook} shows the channel statistics codebook $\mathcal{R}_C$ with $N_{\Delta f} = 62$ and $N_{\Delta t} = 40$, which is designed to cover most of the PDP and Doppler profiles. The pilot configuration $\mathcal{V}$ takes values from the following:% feasible sets:
\begin{enumerate}
\item $\mathcal{P} = \{-10 \text{ dB}, -9 \text{ dB}, -7 \text{ dB}, -5 \text{ dB}, -3 \text{ dB}, 0 \text{ dB}\}$.
\item $\mathcal{D}_f = \{2,4,\cdots, 10, 12 \}$.
\item $\mathcal{D}_t = \{1,2,\cdots, 9,10 \}$.
\end{enumerate}

\begin{figure}[t]
\centering
\includegraphics[width=3.3in]{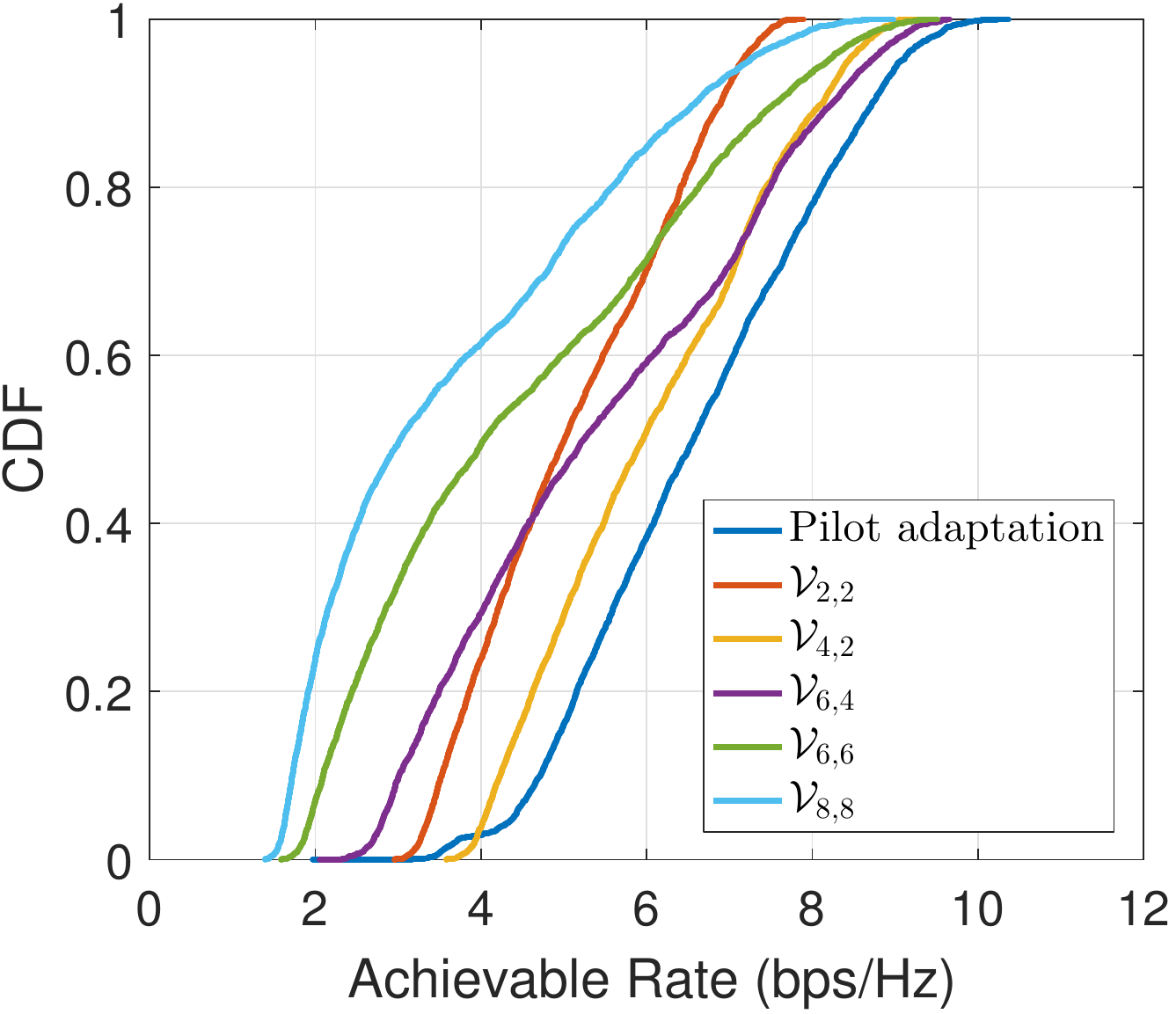}
\caption{CDF comparison of the average achievable rate of pilot adaptation scheme versus fixed pilot configuration schemes. }
\label{Fig4_Achievable_rates}
\end{figure}

Typically, the feasible sets should be chosen such that (a) $\rho$ satisfies the PAPR requirements, (b) $\Delta_p t$ is able to capture the channel variations accurately enough for a large range of vehicular velocities, and (c) $\Delta_p f$ gives reasonably accurate channel estimates for a wide range of multipath environments. In order to estimate the optimal pilot configuration, we use $T_{ofdm}=1500$ OFDM symbols across $N_{sub} = 72$ subcarriers to implement Algorithm \ref{algo_pilot_adapt}. For this case the time duration between the estimation and the use of $\mathcal{V}_o$ is $200 \text{ ms}$, which is less than the stationarity interval of $450 \text{ ms}$. 
%We have chosen $\mathcal{P}, \mathcal{D}_f \text{ and } \mathcal{D}_t$ that is appropriate to the scenario shown in Fig. \ref{Fig3_scenario}. 
\begin{figure}[t]
\centering
\includegraphics[width=3.3in]{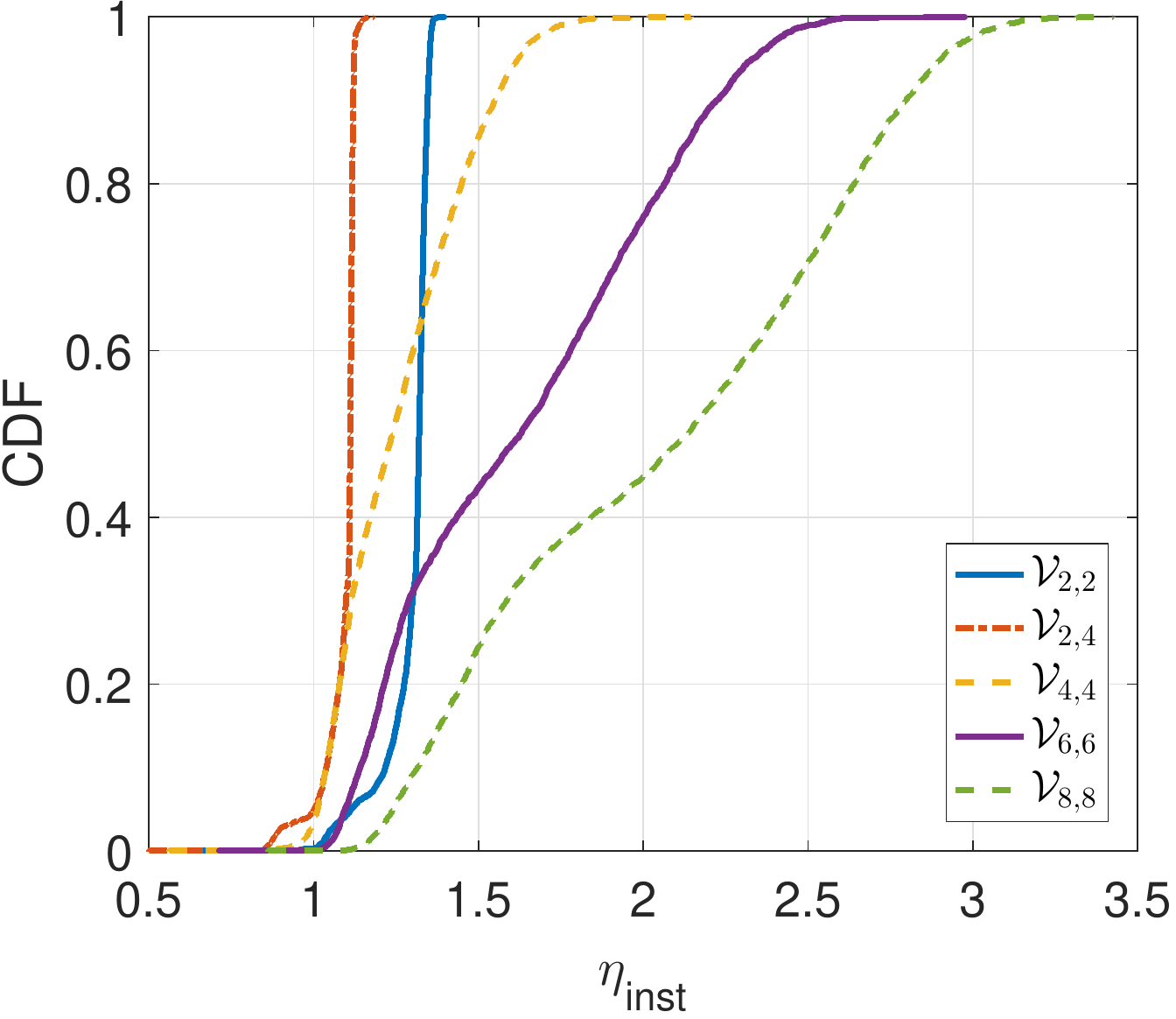}
\caption{CDF comparison of the instantaneous rate gain of pilot adaptation scheme versus fixed pilot configuration schemes.} %$\mathbf{R_{adapt}}$ is the instantaneous rate using the pilot adaptation scheme, while $\mathbf{R_{fixed}}$ is that for the corresponding fixed pilot schemes.}
\label{Fig5_Rate_ratios}
\end{figure} 

Fig. \ref{Fig4_Achievable_rates} shows the cumulative distribution function (CDF) of the achievable rates for all considered pilot configurations. We observe that our proposed adaptive pilot configuration outperforms all the other fixed pilot schemes considered, with the average throughput gain ranging from $9 \%$ to $80 \%$. Fig. \ref{Fig5_Rate_ratios} shows the CDF of the ratio of the instantaneous rates ($\eta_{\mathtt{inst}}$) obtained by the adaptive pilot configuration w.r.t. each considered fixed configuration. Table \ref{Tab3_Gain_comparison} shows the comparison of different percentile values of instantaneous rate gain, with $\Delta \eta_{\mathtt{inst}}^{(x\%)}$ representing the $x$-percentile rate gain. We observe that due to the high Doppler frequencies, the throughput performance deteriorates with higher values of $\Delta_p t$. Even compared to a high pilot density configuration such as $\mathcal{V}_{2,2}$ and $\mathcal{V}_{4,2}$ the proposed pilot adaptation procedure has rate gains ranging from $3.6\%$ to $34.6\%$, demonstrating its efficacy. 

The feedback overhead for explicit and implicit feedback mechanisms is is $\lceil \log_2(6 \times 6 \times 10)\rceil/(1500 \times 71.875\ \mu s) = 83.5 \text{ bps}$ and $\lceil \log_2(6 \times 4)\rceil/(1500 \times 71.875\ \mu s) = 46.4 \text{ bps}$ respectively. Both of these values are negligible compared to the data rates supported by current wireless networks. 
\begin{table}[H]
\renewcommand{\arraystretch}{1.1}
\caption{Instantaneous rate gains of adaptive pilot over fixed pilot configurations}
\label{Tab3_Gain_comparison}
\centering
\begin{tabular}{|c|c|c|c|}
\hline
Scheme & $\Delta \eta_{\mathtt{inst}}^{(10\%)}$ & $\Delta \eta_{\mathtt{inst}}^{(50\%)}$ &  $\Delta \eta_{\mathtt{inst}}^{(90\%)}$ \\
\hline
$\mathcal{V}_{2,2}$ & $21.8\%$ & $32\%$ & $34.6\%$ \\
\hline
$\mathcal{V}_{4,2}$ & $3.8\%$ & $11.1\%$ & $12\%$ \\
\hline
$\mathcal{V}_{6,4}$ & $3.6\%$ & $23.9\%$ & $54.9\%$ \\
\hline
$\mathcal{V}_{6,6}$ & $14\%$ & $62.8\%$ & $122.4\%$ \\
\hline
$\mathcal{V}_{8,8}$ & $31.5\%$ & $113.6\%$ & $179.7\%$ \\
\hline
\end{tabular}
\end{table}
%\vspace{20pt}
\section{Conclusions}\label{Sec5_Conc}
In this paper, we proposed an adaptive pilot configuration mechanism for A2G UAV communications. The receiver estimates channel statistics, maps them to a codebook, finds the optimal parameters and explicitly feeds them back to the transmitter. We compared its throughput performance against several fixed pilot configurations in a scenario where the channel statistics vary over time because of natural variations in the UAV flight environment and speed. We demonstrated average rate gains ranging from $9\%$ to $80\%$, and median instantaneous rate gains ranging from $11\%$ to $114\%$. The signaling feedback overhead for the proposed adaptive pilot configuration method is negligible. It provides a means to accommodate more users for 5G in densely populated UAV networks. For future research, the design of optimal resource allocation algorithms built around this framework is a natural extension to leverage the PHY layer gains of pilot adaptation.
%\section*{Acknowledgements}
%This work was funded in part by the National Science Foundation (NSF) under Grant CNS-1642873. 
\balance
\bibliographystyle{IEEEtran}
\bibliography{references_VTC}
\end{document}